\documentclass[12pt,reqno]{amsart}
\usepackage{amsmath,amssymb,amsfonts,amsthm}
\usepackage[mathscr]{eucal}
\usepackage[russian,english]{babel}
\usepackage{hyperref}
\title{Multiple choice of gauge generators \\ and consistency of interactions}
\author{S.L. Lyakhovich and  A.A. Sharapov}
\address{Physics Faculty, Tomsk State University, Tomsk 634050, Russia}
\email{sll@phys.tsu.ru, sharapov@phys.tsu.ru}

\begin{document}

\maketitle

\begin{abstract}
It is usually assumed  that any consistent interaction either
deforms or retains the gauge symmetries of the corresponding free
theory. We propose a simple model where an obvious irreducible gauge
symmetry does not survive an interaction, while the interaction is
consistent as it preserves the number of physical degrees of
freedom.  The model turns out admitting a less obvious reducible set
of gauge generators which is compatible with the interaction and
smooth in coupling constant. Possible application to gravity models
is discussed.
\end{abstract}

\section{Introduction}
The concept of a consistent interaction first and foremost implies
that the free field equations and the nonlinear ones
describe the same number of physical degrees of freedom. It
also assumes that the Lagrangian and its gauge transformations are
smooth in coupling constants. Proceeding from these assumptions,
one can seek for consistent interactions by adding
vertices to a quadratic Lagrangian and  deforming simultaneously
the gauge symmetry transformations. The deformation technique is
known as the Noether procedure or cohomological perturbation theory,
see \cite{H} for review. If no vertices are compatible with any
deformation of gauge symmetry, this is considered as a no-go theorem
for a consistent interaction. Various no-go results are known for
gravitational interactions in various models, see \cite{HGR},
\cite{JT}, \cite{BBH}, \cite{BDGH} and references therein.

Thus, according to the Noether procedure the interaction  is
considered inadmissible unless a deformation exists for the free
gauge symmetry such that leaves the full Lagrangian invariant. In
the next section, we suggest a simple model that does not correspond
to this wide-spread opinion. In this example, the irreducible (and
most obvious) parametrization of gauge symmetry obstructs
interactions, while the reducible (and less obvious) form of gauge
symmetry is compatible with the interaction. At the free level, both
gauge symmetries are equivalent in the sense that they gauge out the
same number of degrees of freedom, while they are inequivalent with
respect to inclusion of interaction. In Section 3, we discuss a more
complex model of topological gravity where a similar phenomenon can
be expected to appear.

\section{An example of the model with multiple choice of gauge generators}

Consider the following action for the scalar and vector fields in
$2d$ Minkowski space:
\begin{equation}\label{S}
S[\phi, A]=\int d^2x\phi\Big(\partial_\mu A^\mu
+\frac{{g}}{ 2} A_\mu A^\mu\Big)\, .
\end{equation}
The field equations read
\begin{equation}\label{EoM}
\partial_\mu A^\mu +\frac{{g}}{2} A_\mu A^\mu =0 \,,\qquad D^-_\mu\phi=0\,,
\end{equation}
where $D^{\pm}_\mu=\partial_\mu\pm {g} A_\mu$. The parameter ${g}$ plays the
role of the coupling  constant. The commutator of the ``covariant
derivatives'' $D^-_\mu$ gives the ``curvature'' of the vector field
$F=\epsilon^{\mu\nu}\partial_\mu A_\nu$, where
$\epsilon^{\mu\nu}=-\epsilon^{\nu\mu}$ is the $2d$ Levi-Civita symbol.

In the free limit $g \rightarrow 0$, the field equations (\ref{EoM}) have
the obvious general solution\footnote{In this limit (\ref{S}) becomes the action of the 2d abelian $BF$-model, if one identifies the scalar $B$ with the field $\phi$ and expresses the vector field $A^\mu$ in terms of its Hodge-dual $\tilde{A}^\mu=\epsilon^{\mu\nu}A_\nu$. Then, $F=\partial_\mu \tilde{A}^\mu$.}:
\begin{equation}\label{GS}
A^\mu
=\epsilon^{\mu\nu}\partial_\nu\varrho , \qquad \phi  =C\,,
\end{equation}
where $\varrho (x)$ is an arbitrary scalar field, and $C$ is an
arbitrary constant. So, we see that the scalar $\phi$ carries no
local degrees of freedom. The topological mode described by the
constant $C$ is fixed by the boundary conditions. Below we will
always impose zero boundary conditions, so that the unique solution
 will be $\phi=C=0$.

Unlike $\phi$, which is just an auxiliary field vanishing on-shell,
the vector field $A$ may assume arbitrary values at each given instant of time.
This is a direct consequence of the fact that the two
components of $A$ are bound by a single equation. So, if one solves the
field equation for  $\partial_0 A^0$,  the right
hand side will essentially involve the arbitrary function $A^1$,
that makes the solution arbitrary for $A^0$, unless any further
equation is imposed. For example, if we were restricted to the
special solutions with $F=0$, $\varrho$ would not be arbitrary functional parameter
in the general solution (\ref{GS}), rather it would be subject to the
D'Alambert equation $\Box\varrho =0$. Below, we suppose that no
special conditions are imposed on $A$ like that. In particular, we
consider the general solutions with $F(x)\neq 0$ at almost all space-time points.
(This is quite similar to the non-degeneracy assumption for the metric tensor in general relativity.)
Clearly, in the space of all solutions
the vector fields with $F=0$ form a subspace of measure zero.

When $g=0$, the action (\ref{S}) enjoys the irreducible gauge
symmetry
\begin{equation}\label{de}
\delta_\varrho \phi=0\,,\qquad \delta_\varrho A^\mu =\epsilon^{\mu\nu}\partial_\nu\varrho\,,
\end{equation}
with $\varrho $ being the scalar gauge parameter. A simple count
shows that the gauge transformation (\ref{de}) leaves no room for
the local physical degrees of freedom.
So, the theory is topological as might be expected from the analysis of the general solution (\ref{GS}).

With the interaction switched on ($g\neq 0$) the field
$\phi$ is still fixed on the general solution for $A$.
Indeed, the second equation in (\ref{EoM}) has the differential consequence
\begin{equation}
\epsilon^{\mu\nu}D^-_\mu D^-_\nu
\phi= -g F\phi=0\,.
\end{equation}
As the above equation states that the product of two factors
$F\cdot\phi$ vanishes, the dynamics bifurcates into two branches:
either $F=0$ or $\phi=0$. If $F\neq 0$, then $\phi=0$. This branch
has a smooth limit to the case $g=0$, where $\phi$ vanishes, while
$F\neq 0$ for general solutions of this branch. The alternative
option $F=0$ is not smoothly connected with the free case, as it
corresponds to the solution $A_\mu=\partial_\mu \rho$ with $\rho$
subject to the D'Alambert equation $\Box\rho=0$. As we are going to
have the general solution in the free limit, hereafter we opt for
the branch with $\phi = 0$. Then, we still have the first equation
in (\ref{EoM}). It is a single equation imposed on two components of
$A_\mu$:
\begin{equation}\label{Aeq}
\partial_0 A_0+ g/2 (A_0)^2 =\partial_1A_1 + g/2 (A_1)^2 \ .
\end{equation}
The system is under-determined, so the general solution obviously
involves arbitrary function. With arbitrary $A_1$ in the right hand
side, the solution exists for a single unknown function $A_0$. The
general solution for $A_0$ essentially depends on the arbitrary
function $A_1$. From this mere fact one can expect that the
vector field $A$ is pure gauge as it has been in the free theory for
the same reason. Also notice that the field  equation (\ref{Aeq}) has no
differential consequences, in particular, it by no means implies
$F=0$.

Equation (\ref{Aeq}) is smooth in $g$ and describes a system with no
local degrees of freedom for any value of $g$, including $g=0$.
Proceeding from that, one could expect that the gauge symmetry of
the model with $g\neq 0$ is a deformation of the transformation
(\ref{de}) for $g=0$. These expectations, however, do not come true.
The point is that the quadratic vertex $\frac{g}{2} A^2$ in the
equation (\ref{Aeq}) is \textit{not} invariant under the gauge
transformation (\ref{de}) even modulo the free equation,
\begin{equation}\label{drho3}
\delta_\varrho  \left(\frac{g}{2} A^2\right) =  g
A_\mu\epsilon^{\mu\nu}\partial_\nu\varrho \neq 0 \,.
\end{equation}
This means that the gauge symmetry (\ref{de}) can't be deformed to
make it consistent with the quadratic vertex in Eq.
(\ref{Aeq}). If the paradigm of cohomological perturbation theory
was naively applied to this case, it could be interpreted as a no-go
theorem for the interaction.

In our recent paper \cite{LSh}, the existence of a local gauge symmetry
has been proven for any under-determined regular system of $2d$
field equations. In the case under consideration the corresponding gauge transformations read\footnote{The
transformations can be made regular in a vicinity of $F=0$ by
rescaling the gauge parameter: $\varepsilon^\lambda \rightarrow
F^2\varepsilon^\lambda$. Then the special field configurations
($F=0$) are precisely those that are unaffected by the infinitesimal
gauge transformations.}
\begin{equation}\label{GT}
\delta_\varepsilon\phi=0\,,\qquad \delta_\varepsilon A^\mu
={g} \varepsilon^\mu-\epsilon^{\mu\nu}{D^+}_\nu (F^{-1}
D^+_\lambda \varepsilon^\lambda) \,,
\end{equation}
$\varepsilon^\mu $ being an arbitrary vector parameter. Unlike
(\ref{de}), these gauge transformations are reducible. The
corresponding gauge-for-gauge transformations read
\begin{equation}\label{GfGT}
\delta_\varkappa \varepsilon^\mu=\epsilon^{\mu\nu}{D}^+_\nu\varkappa\,,
\end{equation}
where $\varkappa$ is an arbitrary scalar parameter.
The infinitesimal gauge transformations (\ref{GT}) form a closed gauge algebra with the following commutation relations:
\begin{equation}\label{com}
[\delta_{\varepsilon_2},\delta_{\varepsilon_1}]=\delta_{\varepsilon_3}\,,\qquad \varepsilon_3^\mu=\left(\frac{D^+_\lambda \varepsilon^\lambda_2}{F}\right)D^{+\mu}\left(\frac{D^+_\nu \varepsilon^\nu_1}{F}\right)+g\epsilon^{\mu\nu}\varepsilon_{1\nu}\left(\frac{D^+_\lambda \varepsilon^\lambda_2}{F}\right)-(\varepsilon_1 \leftrightarrow \varepsilon_2)\,.
\end{equation}
Notice that the gauge parameter $\varepsilon_3$ is defined here only modulo the reducibility relation (\ref{GfGT}).

Again, a
covariant count of physical degrees of freedom (using, for example,
the general formulae from \cite{KLS}) shows that the transformations
(\ref{GT}),  (\ref{GfGT}) gauge out all the degrees of freedom. So, the
model (\ref{S}) is indeed topological for any $g$.

It is instructive to consider the limit $g\rightarrow 0$ for the
transformation (\ref{GT}, \ref{GfGT}):
\begin{equation}\label{dAdE}
\delta_\varepsilon\phi=0\,,\qquad \delta_\varepsilon
A^\mu=-\epsilon^{\mu\nu}\partial_\nu(F^{-1}\partial_\lambda\varepsilon^\lambda)\,,\qquad
\delta_{\varkappa}\varepsilon^\lambda
=\epsilon^{\lambda\nu}\partial_\nu\varkappa\,.
\end{equation}
As is seen, this reproduces the transformation of the free theory
(\ref{de}) with $\varrho=-F^{-1}\partial_\lambda
\varepsilon^\lambda$.  Since the gauge parameters $\varepsilon^\mu$
enter these transformations through a single function $\varrho$, the
gauge symmetry appears to be reducible. Altogether, the
transformations (\ref{dAdE}) gauge out as many degrees of freedom as
the single gauge transformation (\ref{de}). So, one may regard
(\ref{dAdE}) as a weird form of the ``simplest'', i.e., irreducible
gauge transformation (\ref{de}). We see that the free limit of the
model (\ref{S}) admits a multiple choice for the gauge generators,
including reducible and irreducible options. Both the options
equally well gauge out the degrees of freedom at the free level,
while they are inequivalent from the viewpoint of
interaction\footnote{Strictly speaking relations (\ref{dAdE}) do not
define a complete set of gauge transformations for the free
equations, since the gauge transformation (\ref{de}) can't  be
obtained by specifying the gauge parameter $\varepsilon^\lambda$ in
(\ref{dAdE}). Nonetheless, the set of gauge generators (\ref{dAdE})
is big enough  to gauge out all the degrees of freedom and the
generators form a closed gauge algebra, whose commutation relations
 follow from (\ref{com}) by setting $g=0$.}. The simplest irreducible
choice (\ref{de}) does not survive the interaction, while the less
obvious reducible choice of the gauge transformations (\ref{dAdE})
turns out compatible with the cubic vertex. Some other examples of
multiple-choice of gauge symmetry has been recently noticed in
\cite{FLSh} for free models of various spin fields. As we see here,
the distinctions between the different forms of free gauge
transformations can become crucial at the level of interaction.

This example demonstrates a potential way of bypassing the ``no-go''
theorems for the existence of consistent interactions in various
field-theoretical models. Most of these theorems are deduced from
obstructions to deformation of a particular set of gauge generators.
Similar to the example above, the simplest set of gauge generators
may happen to obstruct any nontrivial deformation, while a less
obvious alternative set can be compatible with reasonable
interactions.

\section{A possible model for topological gravity}
 One can regard the action (\ref{S}) as a pattern for constructing more realistic physical models
 demonstrating the multiple-choice gauge symmetry phenomenon.
 Below, we briefly discuss a theory involving the metric tensor $g$ and the scalar field $\phi$.
 The action reads
\begin{equation}\label{gf}
S[\phi, g]=\int \phi R \,\sqrt{-g} \, d^4x\,,
\end{equation}
where $R$ is the scalar curvature. Rescaling the metric
$g\rightarrow \phi g$, it is even possible to induce a kinetic term
for the scalar field, so that the theory may resemble the
Brans-Dicke gravity \cite{BD}. The equations of motion resulting from
(\ref{gf}) are equivalent to
\begin{equation}\label{R0}
  \qquad R=0\,,\qquad (\nabla_\mu\nabla_\nu-g_{\mu\nu}\Box-R_{\mu\nu})\phi=0\,.
\end{equation}

Consider the linearization of these equations over the background of
$\phi=0$ and and flat metric. The linearized system includes a
single scalar equation for the metric perturbation and the
overdetermined system $(\partial_\mu\partial_\nu - \eta_{\mu\nu}\Box
)\phi=0$ for the perturbation of the scalar field. The general
solution for the latter system is the linear function $\phi=C_\mu
x^\mu +C$, with $C_\mu$ and $C$ being arbitrary constants. Imposing
zero boundary conditions, we get $\phi=0$. So, there is no
propagating degrees of freedom associated with the scalar field.
This fully corresponds to the model of previous section, with the
only difference that the overdetermined system for the scalar field
is now of the second order. One can verify that $\phi$ remains an
auxiliary field on the curved background, though the analysis is
more cumbersome comparing to the previous example. As $\phi$ does
not propagate, the metrics tensor satisfies  the only equation
$R=0$, which is analogous to the equation $\partial_\mu A^\mu
+(g/2)A^2=0$ from the previous section. A regular single equation
essentially involving more than one unknown field should describe a
pure gauge system with no local degrees of freedom. As a result the
model (\ref{gf}) must have a rich gauge symmetry, which is by no
means exhausted by the general coordinate transformations. (The four
parameters of diffeomorphisms are clearly insufficient for gauging
out ten components of the metric tensor.)

Furthermore, the gauge transformations in question can be
reconstructed by the gauge symmetries of the equation $R=0$ alone.
Indeed, if $\delta_\varepsilon g$ is such a symmetry, then
$\delta_\varepsilon R=\hat{A} R$ for some differential operator
$\hat{A}$ depending on $g$, $\varepsilon$,  and their derivatives.
Denoting by $\hat{A}^\ast$ the formal adjoint of the differential
operator $\hat{A}$ with respect to the integration measure
$\sqrt{-g}d^4x$, we can extend the transformation
$\delta_\varepsilon g$ to the gauge invariance of the action
(\ref{gf}) by setting $\delta_\varepsilon \phi=-\hat{A}^\ast \phi$.

Finding a complete set of gauge generators for the action
(\ref{gf}) and their reducibility relations (if any) appears to be a
rather nontrivial problem, which yet to be solved.  The free gauge
transformations, being taken in the most simple form, resist any
deformation to the nonlinear ones, much as it happens in the
example of the previous section. So, there should exist another set
of gauge generators that does not reduce in the flat limit to the
simplest generators of the free theory. We are going to address this
issue elsewhere.

\vspace{0.2cm}

\noindent {\textbf{Acknowledgments.}} We are grateful to D.~Francia
and  M.~Vasiliev  for useful discussions. The work is partially
supported by the Tomsk State University Competitiveness Improvement
Program and the RFBR grant 13-02-00551. A.Sh. appreciates the
financial support from the Dynasty Foundation.

\end{document}